\documentclass[letter]{IEEEtran} 
\usepackage{amsmath}
\usepackage{bm}
\usepackage{amssymb}
\usepackage{graphicx}
\usepackage{color, colortbl} 
\usepackage[dvipsnames,svgnames,x11names]{xcolor}
\usepackage{algpseudocode,algorithm}
\usepackage{hyperref}
\usepackage{enumerate}
\usepackage[caption=false,font=footnotesize]{subfig}
\usepackage{cite}
\usepackage{url,comment}
\usepackage[normalem]{ulem}
\usepackage[shortlabels]{enumitem}
\usepackage{amsthm}
\usepackage{pgfplots}
\usepackage{tikz}
\usepackage{mathtools}
\usepackage{balance}
\usepackage{acronym}
\usepackage{amssymb}

\newcommand{\appropto}{\mathrel{\vcenter{
  \offinterlineskip\halign{\hfil$##$\cr
    \propto\cr\noalign{\kern2pt}\sim\cr\noalign{\kern-2pt}}}}}

\newcommand{\ttt}{\top}
\newcommand{\fim}{\bm{J}}
\acrodef{cdf}[CDF]{cumulative distribution function}
\acrodef{bs}[BS]{base station}
\acrodefplural{bs}[BSs]{base station}
\acrodef{ue}[UE]{user equipment}
\acrodef{los}[LOS]{line-of-sight}
\acrodef{aoa}[AoA]{angle-of-arrival}
\acrodefplural{aoa}[AoA]{angles-of-arrival}
\acrodef{aod}[AoD]{angle-of-departure}
\acrodef{toa}[ToA]{time-of-arrival}
\acrodef{tdoa}[TDoA]{time-difference-of-arrival}
\acrodef{ris}[RIS]{reconfigurable intelligent surface}
\acrodefplural{ris}[RISs]{reconfigurable intelligent surfaces}
\acrodef{tx}[Tx]{transmitter}
\acrodef{rx}[Rx]{receiver}
\acrodefplural{rx}[Rxs]{receivers}
\acrodef{crb}[CRB]{Cram\'er-Rao lower bounds}
\acrodef{rss}[RSS]{received signal strength}
\acrodef{los}[LOS]{line-of-sight}
\acrodef{nlos}[NLOS]{non line-of-sight}
\acrodef{dft}[DFT]{discrete Fourier transform}
\acrodef{fft}[FFT]{fast Fourier transform}
\acrodef{fim}[FIM]{Fisher information matrix}
\acrodef{upa}[UPA]{uniform planar array}
\acrodef{peb}[PEB]{position error bound}
\acrodef{snr}[SNR]{signal-to-noise ratio}
\acrodef{sre}[SRE]{smart radio environment}
\acrodefplural{sre}[SRE]{smart radio environments}
\acrodef{mimo}[MIMO]{multiple-input  multiple-output}
\acrodef{rfid}[RFID]{radio-frequency identification}
\acrodef{ofdm}[OFDM]{orthogonal frequency-division multiplexing}

\bstctlcite{IEEEexample:BSTcontrol}

\title{Semi-Passive 3D Positioning of Multiple RIS-Enabled Users}
\author{
	Kamran~Keykhosravi,~\IEEEmembership{Member,~IEEE,}
	Musa~Furkan~Keskin,~\IEEEmembership{Member,~IEEE,}
	Satyam~Dwivedi,~\IEEEmembership{Member,~IEEE,}
	Gonzalo~Seco-Granados,~\IEEEmembership{Senior~Member,~IEEE,}
	and~Henk~Wymeersch,~\IEEEmembership{Senior~Member,~IEEE}

	\thanks{ This work was supported, in part, by the Swedish Research Council under grant 2018-03701, the Marie Sk\l{}odowska-Curie Individual Fellowships (H2020-MSCA-IF-2019) Grant 888913 (OTFS-RADCOM), 
the Vinnova 5GPOS project under grant 2019-03085,  the EU H2020 RISE-6G project, 	
	the Spanish Ministry of Science, Innovation and Universities under Projects TEC2017-89925-R and by the ICREA Academia Programme. }
	\thanks{K. Keykhosravi, M.~Keskin, and H.~Wymeersch are with the Department of Electrical Engineering at Chalmers University of Technology, 41296 Gothenburg, Sweden (e-mail: kamrank@chalmers.se). 
		}
	\thanks{G. Seco-Granados is with the Department of Telecommunications and Systems Engineering, Universitat Autònoma de Barcelona, 08193 Barcelona,
Spain.}	
\thanks{S. Dwivedi is  with Ericsson Research, Stockholm, Sweden.}	
}

\begin{document}

\maketitle

\begin{abstract}
Reconfigurable intelligent surfaces (RISs) are set to be a revolutionary technology in the 6th generation of wireless systems.  In this work, we study the application of RIS in a multi-user passive localization scenario, where we have one transmitter (Tx) and multiple asynchronous receivers (Rxs) with known locations. We aim to estimate the locations of multiple users equipped with RISs.   The RISs only reflect the signal from the Tx to the Rxs and are not used as active transceivers themselves. Each Rx receives the signal from the Tx (LOS path) and the reflected signal from the RISs (NLOS path).   We show that users' 3D position can be estimated with submeter accuracy in a large area around the transmitter, using the LOS and NLOS time-of-arrival measurements at the  Rxs. We do so, by developing the signal model, deriving the Cram\'er-Rao bounds, and devising an estimator that attains these bounds. Furthermore, by orthogonalizing the RIS phase profiles across different users, we  circumvent  inter-path interference. 
\end{abstract}


\section{Introduction}

Realization of  smart radio environments empowered by \acp{ris}, which enables  ubiquitous communication and radio sensing with high energy and spectrum efficiency, is one of the ambitions of the sixth generation  of  wireless systems \cite{dardari2020communicating}. \ac{ris} consists of a multitude of  unit cells, whose responses to the impinging electromagnetic wave can be controlled, and can thereby improve the quality and coverage of wireless communication  and also enable or improve radio localization \cite{ConvergentCommunication,wymeersch2019radio}. In addition to these benefits, \ac{ris}s are semi-passive devices with low cost, which make them ideal to be mounted on surfaces (e.g., walls) as well as moving objects (e.g., vehicles).  

\begin{figure}
    \centering
    \begin{tikzpicture}
    \node  [anchor=mid] (pic) {\includegraphics[width=\linewidth]{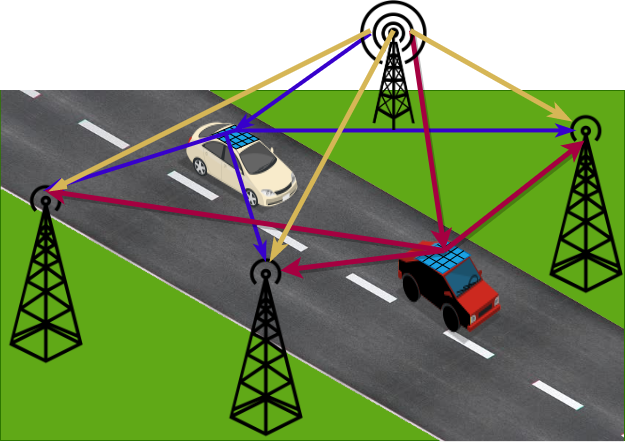}};
    \node (c1) at (1.8, 6) {\footnotesize Tx};
    \end{tikzpicture}
    \caption{A schematic of the system  for two UEs ($N=2$) and three Rxs ($M=3$).}
    \label{fig:system}
\end{figure}

Radio localization has attracted  increasing attention in recent years as technologies such as millimeter wave, \ac{mimo}, and \ac{ris} enable high-accuracy positioning of users based on the \ac{toa} and angles-of-arrival and -departure measurements \cite{Shahmansoori18TWC}. Considering the nature of the user, localization techniques can be categorized into active and passive methods. While with the former case the user transmits or receives signals (e.g., in \cite{Shahmansoori18TWC}), in the latter case, the user only reflects or scatters the signals from a \ac{tx} (see e.g., \cite{eigenvector}). Many studies have been conducted on passive localization based on a variety of approaches such as \ac{rfid} \cite{rfidMag,VehichlesOnRfid}, signal eigenvectors \cite{eigenvector}, \ac{rss} \cite{ruan2014tagtrack}, and \ac{toa}-based passive positioning \cite{shen2012accurate,wang2014toa,ExpectationYuan,decarli2013novel,delayDoppler_passive_TSP_2016}. In the latter case, which is the focus of this letter, the user location is estimated based on the received signal \ac{toa} at multiple  \acp{rx}. This topic has been studied in two-dimensional space under the assumption of synchronous \cite{shen2012accurate}, quasi-synchronous \cite{wang2014toa}, and asynchronous networks \cite{ExpectationYuan}.  In \cite{decarli2013novel}, the authors study the 2D localization performance of a joint radar and \ac{rfid} system. Moreover, bistatic \ac{toa} estimation has been investigated in passive sensing systems that employ the signals transmitted by illuminators of opportunity (IO) \cite{delayDoppler_passive_TSP_2016}. To the best our knowledge, this is the first paper on passive localization of \ac{ris}-enabled users.

In this work, we investigate the multi-user 3D passive positioning problem, employing one \ac{tx} and multiple \acp{rx}, where each user is equipped with an \ac{ris} (see Fig.\,\ref{fig:system}). We propose a low-complexity positioning algorithm, which utilizes orthogonal sequences in the design of \ac{ris} phase profiles. By employing  the orthogonality property of the received signal, the algorithm can resolve  multipath interference and the data association problem. In other words, it can decompose the received signal at each \ac{rx} into the \ac{los} component and the signals reflected from each \ac{ue}. Thereafter, the \ac{toa}  can be readily estimated at each \ac{rx} for each of the multipath components, which enables localization of the \ac{ue}s. Finally, we evaluate the localization error of the proposed method and show that it reaches the theoretical \ac{crb}.

\subsection{Notation}
Vectors, which are columns, are shown by bold lower-case letters and matrices by bold upper-case ones. The element at the $i$th row and the $j$th column of the matrix $\bm{A}$ is shown as $[\bm{A}]_{i,j}$.  The sets $\mathbb{C}$ and $\mathbb{T}$ represents the set of complex numbers and all the complex numbers with unit magnitude, respectively. The vector $\mathbf{1}$ indicates the all-ones vector and the operator $\circ$ specifies the element-wise multiplication.

\section{System Model}

\subsection{Signal Model}
We consider one Tx (a \ac{bs}) with \emph{known} location $\bm{p}_0$ and $M$ \ac{rx}s (\acp{bs} or road-side units) with \emph{known} locations $\bm{p}_1,\ldots, \bm{p}_M$,  as well as $N$ \ac{ue}s with \emph{unknown} locations  $\bm{x}_1,\ldots,\bm{x}_N$. Each of the \ac{ue}s is equipped with an \ac{ris}, while the Tx and  Rxs have a single antenna (the analysis and method also applies to \ac{tx} and \acp{rx} with multiple antennas). The \ac{rx}s are not synchronized with the \ac{bs} and have \emph{unknown} clock biases $B_1,\dots, B_M$. Each \ac{rx} receives the signal directly from the \ac{tx}, which is the \ac{los} path, and also the reflected (\ac{tx}--\ac{ris}--\ac{rx}) signals from \acp{ris}, which is the \ac{nlos} path).
We consider the transmission of $T$ \ac{ofdm} symbols with $K$ subcarriers during each localization occasion, where we assume that  $T$ is sufficiently small, so that \ac{ue} mobility can be ignored, i.e., the user movement during the transmission is much less than (e.g., $10\%$) the wavelength.

The signal received at the $m$th \ac{rx}, after cyclic prefix removal and \ac{fft}, can be represented by the matrix $\bm{Y}_m\in\mathbb{C}^{K\times T}$. Assuming constant pilot transmission over all subcarriers, we have (see for example \cite{keykhosravi2020siso})
\begin{align}
    \bm{Y}_m =   \sqrt{E_{\mathrm{s}}}\sum_{n=0}^{N} \bm{d}(\tau_{n,m}) \bm{\alpha}_{n,m}^\top + \bm{W}_m,\label{eq:obs1}
\end{align}
where ${E_{\mathrm{s}}}$ is the symbol energy and 
\begin{align}
    \bm{d}(\tau) & = [1, e^{\jmath 2 \pi \Delta f \tau}, \dots, e^{\jmath 2 \pi (K-1) \Delta f \tau}]^\top
\end{align}
represents the phase offset produced by the delay on each subcarrier, where $\Delta f$ is the subcarrier spacing. 
For the \ac{los} path ($n=0$), the delay is $\tau_{0,m}=\Vert\bm{p}_0- \bm{p}_m\Vert/c + B_m/c $, in which the distance $\Vert\bm{p}_0- \bm{p}_m\Vert$ is known and $c$ is the speed of light. For the reflected paths ($n>0$), the delay is %
\begin{align}
    \tau_{n,m} = \frac{\Vert\bm{p}_0- \bm{x}_n\Vert+\Vert\bm{x}_n- \bm{p}_m\Vert+ B_m}{c}.
\end{align}
The vector $\bm{\alpha}_{n,m}\in \mathbb{C}^{T \times 1}$ represents the complex gain of different paths. For $n=0$ (LOS) $\bm{\alpha}_{0,m} = \alpha_{0,m} \bm{1}_T$, where $\alpha_{0,m}$ indicates the \ac{los}  gain. For $n \neq 0$, we have
\begin{align}
    [\bm{\alpha}_{n,m}]_t = \gamma_{n,0}\gamma_{n,m}\bm{a}(\bm{\theta}_{n,m})^\top \bm{\Omega}_{n}[t]\bm{a}(\bm{\phi}_{n,0})
\end{align}
in which $\gamma_{n,0}$ is the complex channel gain from the transmitter to \ac{ue} $n$ and $\gamma_{n,m}$ is the complex channel gain from \ac{ue} $n$ to receiver $m$. 
The noise matrix is represented by $\bm{W}_m\in\mathbb{C}^{K\times T}$, which has i.i.d circularly-symmetric Gaussian elements and variance $N_0$.
Moreover, $\bm{a}(\bm{\theta}_{n,m})$  is the steering vector as a function of the \ac{aod} ($\bm{\theta}_{n,m}$) from the $n$th \ac{ue} to the $m$th \ac{rx}, measured in the unknown frame of reference of \ac{ue} $n$. Let $\bm{R}_n$ indicate the unknown rotation matrix mapping  the global frame of reference to  the coordinate system associated with the $n$th \ac{ris}.   Then \ac{aod} $\bm{\theta}_{n,m}$ represents the angle in the direction of vector $\bm{w}_{n,m} = \bm{R}_n(\bm{p}_m - \bm{x}_n)$, i.e., $[\bm{\theta}_{n,m}]_{\mathrm{az}} = \mathrm{atan2}([\bm{w}_{n,m}]_2,[\bm{w}_{n,m}]_1)$ and $[\bm{\theta}_{n,m}]_{\mathrm{el}} = \mathrm{acos}([\bm{w}_{n,m}]_3/\Vert\bm{w}_{n,m}\Vert)$. Similarly, $\bm{a}(\bm{\phi}_{0,n})$ indicates the $n$th \ac{ris} steering vector at \ac{aoa} ($\bm{\phi}_{0,n}$) from the \ac{tx} to the $n$th \ac{rx}, which is the  angle associated with the vector $\bm{v}_{n, 0} = \bm{R}_n(\bm{p}_0-\bm{x}_n)$. The steering vector at angle $\bm{\psi}$ for an  \ac{ris} with $W_{\mathrm{r}}\times W_{\mathrm{c}}$ elements on the $x-y$ plane in the \ac{ris} coordinate system and distance $d$ between adjacent elements is $\bm{a}(\bm{\psi}) = \bm{a}_{\mathrm{r}}(\bm{\psi})\otimes \bm{a}_{\mathrm{c}}(\bm{\psi})$, where
\begin{align}
    \bm{a}_{\mathrm{r}}(\bm{\psi}) &= e^{\jmath\beta_{\mathrm{r}}}[1, e^{\jmath d [\bm{k}(\bm{\psi})]_1}, \dots, e^{\jmath (W_{\mathrm{r}}-1) d [\bm{k}(\bm{\psi})]_1} ]^\top\\
    \bm{a}_{\mathrm{c}}(\bm{\psi}) &= e^{\jmath\beta_{\mathrm{c}}}[1, e^{\jmath d [\bm{k}(\bm{\psi})]_2}, \dots, e^{\jmath (W_{\mathrm{c}}-1) d [\bm{k}(\bm{\psi})]_2} ]^\top
\end{align}
where $\beta_{\mathrm{r}} = -(W_{\mathrm{r}}-1)d [\bm{k}(\bm{\psi})]_1/2$ and $\beta_{\mathrm{c}} = -(W_{\mathrm{c}}-1)d [\bm{k}(\bm{\psi})]_2/2$ and
\begin{align}
    \bm{k}(\bm{\psi}) = \frac{2\pi}{\lambda}[\sin\psi_{\mathrm{el}}\cos\psi_{\mathrm{az}},\sin\psi_{\mathrm{el}}\sin\psi_{\mathrm{az}},\cos\psi_{\mathrm{el}}]^\top
\end{align}
is the wavenumber vector. The elevation angle is measured from the $z$ axis and the azimuth angle in the $x-y$ plane from the $x$ axis. 
 Finally,  $\bm{\Omega}_{n}[t] \in \mathbb{C}^{W\times W}$, where $W = W_{\mathrm{r}}W_{\mathrm{c}}$, is a  diagonal matrix that represents the phase profile  of \ac{ris} $n$ as a function of time $t$.

\subsection{Problem formulation} \label{sec:problemFormul}
Our goal is to estimate the locations of the $N$ \ac{ue}s, $\bm{x}_1,\ldots,\bm{x}_N$ from  $\{Y_m\}_{m=1}^{M}$ in \eqref{eq:obs1}. To do this, we propose the following approach.
\begin{itemize}
    \item To estimate at \ac{rx} $m$, the $N+1$ \ac{toa}s $\tau_{n,m}$. For this, we use the design freedom of the \ac{ris} in terms of $\bm{\Omega}_{n}[t]$ to avoid interference from different paths. 
    \item To compute \ac{tdoa} measurements at each of the $M$ \ac{rx}s and process them jointly  to localize all users. 
\end{itemize}

\section{Methodology}
In this section, we  address the two steps mentioned in Section\,\ref{sec:problemFormul}. We first introduce a special RIS phase profile design in Section\,\ref{sec:risPhase} that allows us to decouple the received signals at each \ac{rx}.  Then based on the received signals we estimate \ac{toa}s in Section\,\ref{sec:toaEst}. Finally, in Section\,\ref{sec:posEst}, we use the \ac{toa}s to estimate the position of the \ac{ue}s.  

\subsection{RIS phase profile design}\label{sec:risPhase}
We design the phase profile of each \ac{ris}  to avoid the interference between different signal paths.  To do so, for  \ac{ue} $n$, we set the \ac{ris} profile $\bm{\Omega}_{n}[t]$ to be the product between a constant diagonal matrix $\bm{\Omega}_{n}\in \mathbb{C}^{W\times W}$ and a  time-varying  scalar $[\bm{\omega}_n]_t \in \mathbb{T}$, i.e., $\bm{\Omega}_{n}[t]=[\bm{\omega}_n]_t\bm{\Omega}_{n}$. We also define $\bm{\omega}_0=\bm{1}_T$, without loss of generality. As will be shown in Section~\ref{sec:toaEst}, we can avoid  inter-path interference if the vectors $\bm{\omega}_n \in \mathbb{T}^{T\times 1}$ for $n = 0, 1, \dots, N$  form an orthogonal set, i.e.,  
\begin{align}
    \bm{\omega}^H_n \bm{\omega}_{n'} & =\begin{cases}T  &\mbox{if } n=n' \\
0 & \mbox{otherwise.}  \end{cases}\label{eq:prop2}
\end{align}
Therefore, one should set the the number of transmission $T$ higher than $N$ to be able to select $N+1$ orthogonal vectors $\{\bm{\omega}_{n}\}_{n=0}^{N}$.
We choose the vector $\bm{\omega}_n$ to be the $n$th column of the $T\times T$ \ac{dft} matrix $\bm{F}$ with elements 
\begin{align}\label{eq:ifft}
    [\bm{F}]_{\ell,m} = e^{-2\jmath \pi \ell m/T}.
\end{align}
We assume infinite resolution for the phase shifts of the \ac{ris} unit cells. However, in practice, the resolution might be limited to few bits, in which case the selection of $\{\bm{\omega}_{n}\}_{n=0}^{N}$ should be adapted to this restrictions. This problem may be solved using prior works on code-division multiple access systems (see e.g., \cite{fan2004spreading}). We leave the study of this problem to future works.  

In terms of the constant part $\bm{\Omega}_{n}$, since  we do not assume any prior knowledge of the user location, we set this part randomly. However, if  an initial estimation of the user location and orientation is available, one can design $\bm{\Omega}_{n}$ to obtain a higher \ac{snr} at the \ac{rx}s.

\subsection{ToA estimation at Tx $m$}\label{sec:toaEst}
In order to estimate $\tau_{n,m}$ at \ac{rx} $m$ and for $n = 0\dots N$, we make use of \eqref{eq:prop2} by computing 
\begin{align}
    \bm{r}_{n,m}&=\frac{1}{T}\bm{Y}_m\bm{\omega}_n^*\\
   & =\sqrt{E_{\mathrm{s}}}\beta_{n,m} \bm{d}(\tau_{n,m}) + \bm{z}_{n,m}\label{eq:rnm}
\end{align}
where $\bm{z}_{n,m} = \bm{W}_m\bm{\omega}_n^*/T$ and it can be shown that $\mathbb{E}\{ \bm{z}_{n,m} \bm{z}^H_{n,m}\}=N_0/T \bm{I}$. Also,
\begin{align}
    \beta_{n,m}=\begin{cases}
    \alpha_{0,m}& \text{if } n= 0\\
    \gamma_{0,n}\gamma_{m,n}\bm{a}(\bm{\theta}_{m,n})^\top \bm{\Omega}_{n}\bm{a}(\bm{\phi}_{0,n})              & \text{otherwise}.
    \end{cases}
\end{align}

From this observation, we can easily determine $\tau_{n,m}$ using standard methods.  Here, we use  \ac{fft} with a refinement step based on quasi-Newton method \cite{keykhosravi2020siso}. We explain this method in brief for completeness. Upon receiving vector $\bm{r}_{n,m}$, we calculate  $\bm{r}_{n,m}(\delta)=\bm{r}_{n,m}\circ \bm{d}(\delta)$, which mimics a delayed version of $\bm{r}_{n,m}$ in the frequency domain. Let $\bm{b}_{n,m}(\delta)$ be the $F$-point \ac{fft} of the vector $\bm{r}_{n,m}(\delta)$, where $F$ is a design parameter. Then we estimate $\tau_{n,m}$ as  $\hat{\tau}_{n,m} =\Tilde{k}/(F\Delta f) - \Tilde{\delta}$, where $[\Tilde{\delta},\Tilde{k}] = \arg\max_{k,\delta} \vert[\bm{b}_{n,m}(\delta)]_k\vert$ and $\delta\in[0,1/(F\Delta f)]$. This 2D optimization can be divided to two 1D ones \cite{keykhosravi2020siso}.

\subsection{Estimating the position of user $n$} \label{sec:posEst}
We compute the \ac{tdoa} measurements 
\begin{align}\label{eq:deltamn}
    \Delta_{n,m}& =c(\hat{\tau}_{n,m}-\hat{\tau}_{0,m}) +\Vert\bm{p}_0- \bm{p}_m\Vert\\
    & =\Vert\bm{p}_0- \bm{x}_n\Vert+\Vert\bm{x}_n- \bm{p}_m\Vert + w_{n,m},\label{eq:delta_mn}
\end{align}
where we use the \ac{los} paths as  references to remove the clock biases $B_m$. Eq.~\eqref{eq:delta_mn} defines an ellipsoid in 3D
with foci $\bm{p}_0$ and $\bm{p}_m$. For each \ac{ue} $n$, we aggregate all the measurements in $\bm{\Delta}_n = [\Delta_{n,1}, \dots, \Delta_{n,M}]^\top$ across different \ac{rx}s and the corresponding noises in $\bm{w}_n $, where we model $\bm{w}_n\sim \mathcal{N}(\bm{0},\bm{\Sigma}_n)$. Note that $\bm{\Sigma}_n$ is a diagonal matrix (which is different from the standard \ac{tdoa} localization system), since the noises at different \ac{rx}s are uncorrelated. The elements of $\bm{\Sigma}_n$ can be estimated using the \ac{crb}  for $\tau_{n,m}$, which can be calculated based on \cite[Chapter~3]{kay1993fundamentals} as  
\begin{align}\label{eq:crb_tau}
    \mathrm{E}[|\tau_{n,m}-\hat\tau_{n,m}|^2]\geq \frac{6N_0}{K (K^2-1)TE_{\mathrm{s}}|2\pi \Delta f  \beta_{n,m}|^2}.
\end{align}
Then based on \eqref{eq:deltamn}, the covariance matrix $\bm{\Sigma}_n$ can be calculated  as
\begin{align}
    [\bm{\Sigma}_{n}]_{m,m} = c^2 \left(\mathrm{E}[|\tau_{n,m}-\hat\tau_{n,m}|^2]+\mathrm{E}[|\tau_{0,m}-\hat\tau_{0,m}|^2]\right)
\end{align}
where in order to calculate \eqref{eq:crb_tau} we estimate $|\beta_{n,m}|$ based on \eqref{eq:rnm} as 
\begin{align}
    |\beta_{n,m}| = \left\vert\frac{\bm{d}(\hat{\tau}_{n,m})^\ttt\bm{r}_{n,m}}{\sqrt{E_{\mathrm{s}}}\bm{d}(\hat{\tau}_{n,m})^\ttt\bm{d}(\hat{\tau}_{n,m})}\right\vert.
\end{align}
We introduce $\bm{\Delta}_n = \bm{h}(\bm{x}_n)+ \bm{w}_n$, where $[\bm{h}(\bm{x})]_m = \Vert\bm{p}_0- \bm{x}\Vert+\Vert\bm{x} - \bm{p}_m\Vert$. 
We thus find the \ac{ue} location estimate as 
\begin{align}\label{eq:ML}
    \hat{\bm{x}}_n = \arg \min_{\bm{x}_n} (\bm{\Delta}_n - \bm{h}(\bm{x}_n))^\ttt \bm{\Sigma}_n^{-1} (\bm{\Delta}_n - \bm{h}(\bm{x}_n))
\end{align}
which can be solved via gradient descent algorithm, starting from an initial guess. We now propose a method to find such an initial guess.

\begin{figure*}
    \centering
    \includegraphics{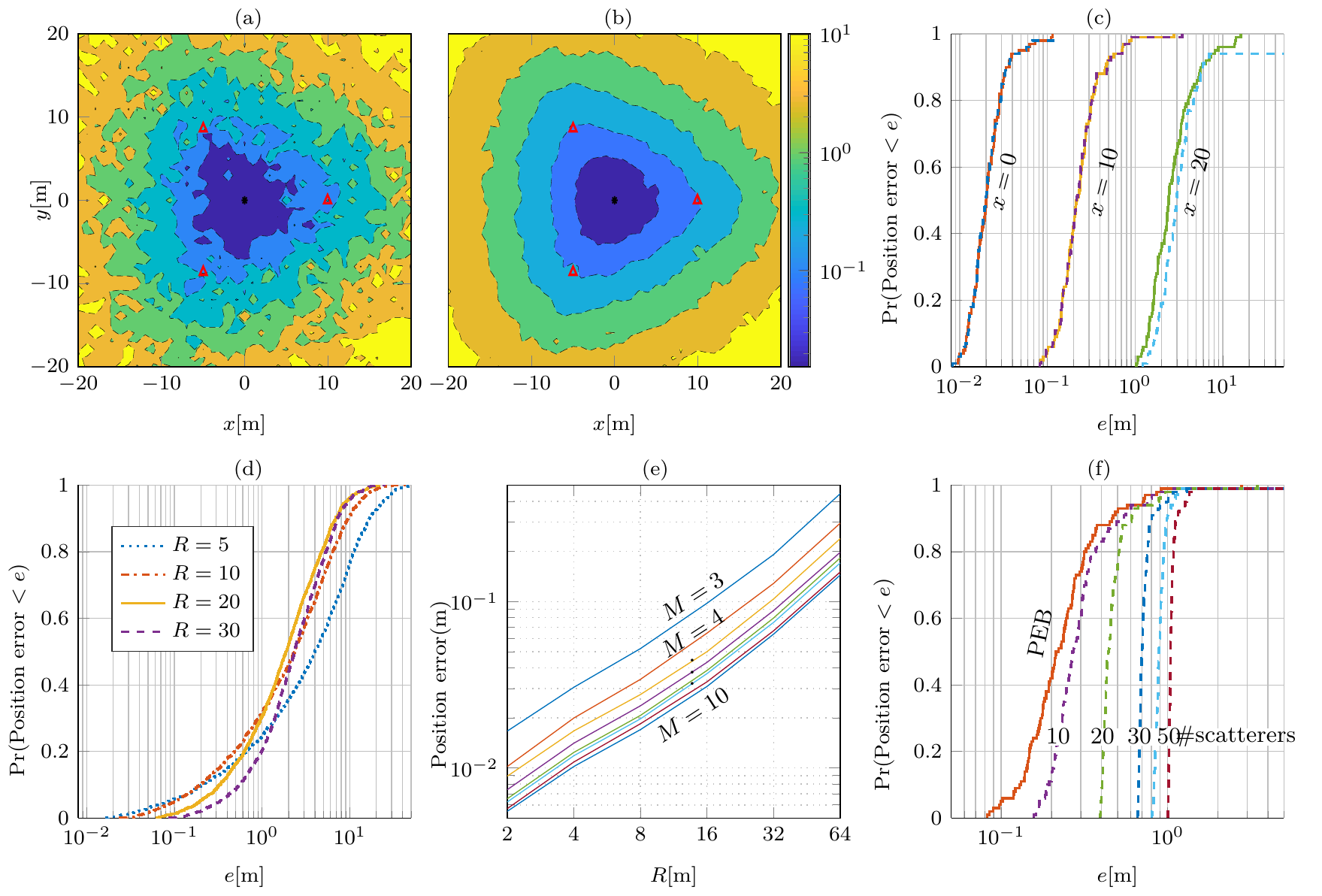}
    \caption{The position error (in meters) for a  system with one Tx at the origin (marked by a black circle), $M=3$ Rxs (marked by red triangles) uniformly located on a circle with radius $R=10$ on the plane $z=1$, and a user located on $[x,y,-3]$:  (a)  \ac{peb} for one random RIS configurations; (b) the average of \ac{peb} over $100$ random RIS configurations; (c) the CDF of \ac{peb} (solid lines) and  estimation error (dashed lines) for $100$ random RIS configurations at $[x,0,-3]$ $x\in\{0,10,20\}$; (d) the CDF of average \ac{peb} for $41\times 41$ equispaced user locations shown in subfigure (a) with $R\in\{5,10,20,30\}$; (e)  average \ac{peb} at $[0,0,-3]$ for $M=3\dots 10$; (f) average \ac{peb} (solid line) and the estimation error (dashed lines) for $100$ random RIS configurations at $[10,0,-3]$ in the presence of $10, 20, 30, 40,$ and $50$ scatterers.    }
    \label{fig:results}
\end{figure*}

Without loss of generality, we set $\bm{p}_0=\bm{0}$. In the absence of noise and based on \eqref{eq:delta_mn}, we have that $    (\Delta_{n,m}-\Vert \bm{x}_n\Vert)^2 =\Vert \bm{x}_n-\bm{p}_m\Vert^2$, which leads to
\begin{align}
    \bm{p}_m^\ttt\bm{x}-\Delta_{n,m}\Vert \bm{x}_n\Vert& = \frac{1}{2}(\Vert \bm{p}_m\Vert^2-\Delta_{n,m}^2). \label{eq:linearSystem}
\end{align}
We can rewrite \eqref{eq:linearSystem} in the matrix form as  $\bm{P} \bm{x}_n = \bm{z}_n+ \bm{\Delta}_n \Vert \bm{x}_n\Vert$, where $\bm{P}=[\bm{p}_1,\bm{p}_2,\dots, \bm{p}_M]^\ttt$, $\bm{z} = 0.5[\Vert \bm{p}_1\Vert^2-\Delta_{n,1}^2, \dots, \Vert \bm{p}_M\Vert^2-\Delta_{n,M}^2]^\ttt$.
Then the $n$th user position can be estimated as \cite{malanowski2009algorithm}
\begin{align}
    \hat{\bm{x}}_n &= \bm{a}_n + \bm{b}_n \Vert \hat{\bm{x}}_n\Vert
\end{align}
where $\bm{a}_n = (\bm{P}^\ttt \bm{P})^{-1}\bm{P}^\ttt \bm{z}_n$, $\bm{b}_n = (\bm{P}^\ttt \bm{P})^{-1}\bm{P}^\ttt  \bm{\Delta}$, and 
\begin{align}
    \Vert \hat{\bm{x}}_n\Vert& = \frac{-\bm{a}_n^\ttt\bm{b}_n\pm\sqrt{(\bm{a}_n^\ttt\bm{b}_n)^2-\Vert\bm{a}_n\Vert^2(\Vert\bm{b}_n\Vert^2-1)}}{\Vert\bm{b}_n\Vert^2-1}.\label{eq:estAbsX}
\end{align}
If \eqref{eq:estAbsX} yields two viable solutions, one can insert both solutions to the negative log-likelihood function, which is the objective function in \eqref{eq:ML}. If the outcome for one of the solutions is much smaller than the other one, then it  should be used as the initial guess for the $n$th user position. However, if both outcomes are small, this indicates that the $M$ ellipsoids in \eqref{eq:delta_mn} intersect in two distinct points. In such a case some prior knowledge (e.g.,  the user  is located in a given region in relation to the  \ac{rx}s) should be used to localize the user.

\section{Simulation Results}

In this section, we evaluate the estimation error of the user position and compare it to the theoretical \ac{peb}. The \ac{ris} is a $256\times 256$ \ac{upa}.   The clock biases $B_m$ are selected uniformly in the interval $[0,1/\Delta f)$. Since there is no interference between the LOS path and the \ac{nlos} paths from different users, the performance of the estimator for  each user is independent of the number of users $N$ (as long as $T>N$) and therefore, we set $N=1$.  For the \ac{los} path, the channel gain $\alpha_{0,m}$ is calculated based on Friis’ formula assuming unit directivity for \ac{tx} and \ac{rx}s. For the \ac{nlos} path the channel gain is calculated as \cite[Eq. (21)--(23)]{ellingson2019path}
\begin{align}
    \gamma_{n,0}\gamma_{n,m} = \frac{\lambda^2(\cos(\theta_{n,m})\cos(\phi_{n,0}))^{0.285}}{16 \pi\Vert \bm{p}_{0} - \bm{x}_{n}  \Vert\Vert \bm{p}_{m} - \bm{x}_{n}  \Vert}.
\end{align}
All the rotational angles corresponding to the user orientation is set to  zero ($\bm{R}_n = \bm{I}_3, \ \forall n$). The diagonal elements of $\bm{\Omega}_{n}$ are drawn randomly and independently from the unit circle. The presented results are obtained by averaging over $10,000$ random realization of RIS phase profiles ($100$) and noise ($100$ for each RIS phase profile). In our estimator, we use the prior knowledge that the \ac{ue} is below the \ac{rx}s to resolve the sign ambiguity in \eqref{eq:estAbsX}. The rest of the system parameters are represented in Table\,\ref{table:par}.

\begin{table}[!t]
\vspace{.1cm}
	\caption{Parameters used in the simulation.}
	\label{table:par}
	\centering
	\begin{tabular}{l l l }
		\hline
		\hline
		Parameter&Symbol& Value\\
		\hline
		Wavelength & $\lambda$ & $1 \ {\mathrm{cm}}$\\
		RIS element spacing & $d$ &$0.5 \ {\mathrm{cm}}$\\
		Light speed & $c$ & $3\times 10^8 \ \mathrm{m/s}$\\
		Number of subcarriers & $K$ & $100$\\
		Subcarrier bandwidth & $\Delta f$ & $120 \ \mathrm{kHz}$\\
		Number of transmissions & $T$ & $32$\\
		Transmission Power &$K E_\mathrm{s}\Delta f$ & $25 \ \mathrm{dBm}$\\
		Noise PSD & $N_0$ & $-174 \ \mathrm{dBm/Hz}$\\
		UE's Noise figure& $n_f$ & $5 \ \mathrm{dB}$\\
		FFT dimensions & $F$ & $1024$\\
		\hline
		\hline
	\end{tabular}
\end{table}

In Fig.\,\ref{fig:results} the position error has been analyzed for a system with the \ac{tx} at the origin, $M=3$ Rxs located on a circle with radius $R=3\ \mathrm{m}$ on the plane $z=1\ \mathrm{m}$, and a user located on $z=-3\ \mathrm{m}$. Fig.\,\ref{fig:results}(a)  illustrates the \ac{peb} for one realization of the \ac{ris} phase profile while Fig.\,\ref{fig:results}(b) does so for the average of the \ac{peb} over $100$ random \ac{ris} configurations. It is evident that submeter localization accuracy can be attained  in a large area around the \ac{tx}. From Fig.\,\ref{fig:results}(b) one can see that  the average \ac{peb} gradually and symmetrically increases with the distance from the \ac{tx}. In general, the same behavior can be observed also in Fig.\,\ref{fig:results}(a), however, the increase in \ac{peb} is not smooth and symmetrical, which is due to different \ac{ris} reflection gains in different directions for a  random phase profile. Fig.\,\ref{fig:results}(c) represents the \ac{cdf} of the \ac{peb} and the estimation error for $100$ \ac{ris} configurations at three points. It can be seen that the presented estimator tightly attains the the \ac{peb} as long as the \ac{peb} is less than $8$ meters. 

In Fig.\,\ref{fig:results}(d) we study the effect of \ac{rx} placement on the average \ac{peb}. To do so, we evaluate the \ac{cdf} of the average \ac{peb} for a $41\time41$ grid of \ac{ue} locations over the area shown in  Fig.\,\ref{fig:results}(a). We consider four different values for the horizontal distance from \ac{rx} to \ac{tx} ($R$). It can be seen that for the majority of \ac{ue} locations $R=20\ \mathrm{m}$ obtains superior position accuracy. This indicates that  \ac{rx}s should be placed close to the edge of the region of interest to improve the (worst-case)  localization accuracy.
 
In Fig.\,\ref{fig:results}(e) the \ac{peb} at a UE position equal to $[0,0,-3]$ is calculated for different \ac{rx} numbers ($M$) and positions ($R$) of receivers. It can be seen that \ac{peb} increases with $R$ linearly, which is due to the quadratic decrease of \ac{snr}. Furthermore, it is evident that the improvement in \ac{peb} achieved by increasing the number of \ac{rx}s is noticeable only for low values of $M$.

Fig.\,\ref{fig:results}(f) illustrates the \ac{peb} and the estimation error at the UE position $[10,0,-3]$ for $100$ realizations of \ac{ris} configuration in the presence of additional scatterers. The scatterers are placed randomly one meter below the \ac{ue}s and within $10\,\mathrm{m}$ radius of the point $[0,0,-4]$. Since the scatterers are below the RIS, they only scatter the signal coming directly from the Tx onto them and not the reflected signals from the \acp{ris}.
The channel gain for the scattered signal is calculated based on the radar range equation by assuming radar cross section of $0.1\,\mathrm{m}^2$.  The interference from the scatterers  deteriorates our estimation accuracy of the \ac{los} delay $\hat{\tau}_{0,m}$, however, it does not affect that of the \ac{nlos} delay $\hat{\tau}_{n,m}$ ($n>0$). This is because the interference from scatterers cancels out upon calculating $\bm{r}_{n,m}=(1/T)\bm{Y}_m\bm{\omega}_n^*$  since we have $\bm{1}_T^\ttt \bm{\omega}_n^* = 0, \forall n>0$. With a large number of scatterers, the position error is mainly affected by the error in \ac{los} \ac{toa} estimation and therefore is  predominantly independent of RIS phase profile, hence the sharp transition of CDF. Finally, we note that the estimator can perform properly even in the presence of a large number of scatterers and obtain submeter localization accuracy.

\vspace{-.1cm}
\section{Conclusion}
We considered a multi-user RIS-enabled  localization problem, where the users' position in 3D was estimated by calculating the \ac{toa} of the \ac{los}  and \ac{nlos} paths at multiple receivers. The considered scenario can be categorized as a passive localization problem since the users do not generate transmitted signal or process received signals, but only reflect signals, based on which their positions are obtained. Nonetheless, it should be noted that RISs are not completely passive as they require some source of energy to reconfigure.  We showed that by dividing the \ac{ris} phase profile to  constant and time-varying parts and selecting the time-varying one based on orthogonal sequences, the  interference between  all the reflected NLOS signals among themselves and with the LOS paths can be avoided.  In future work we aim to optimize the constant part of the RIS phase profile to improve the \ac{snr} of the \ac{nlos} path and achieve better localization accuracy. Extending the work to account for the limitations in \ac{ris} phase resolution and \ac{ris} synchronization is also an interesting future direction.

\vspace{-.1cm}
\section*{Acknowledgment}
The authors gratefully acknowledge  contributions of Jonas~Medbo in seeding the idea of this article.


\begin{thebibliography}{10}
	\providecommand{\url}[1]{#1}
	\csname url@samestyle\endcsname
	\providecommand{\newblock}{\relax}
	\providecommand{\bibinfo}[2]{#2}
	\providecommand{\BIBentrySTDinterwordspacing}{\spaceskip=0pt\relax}
	\providecommand{\BIBentryALTinterwordstretchfactor}{4}
	\providecommand{\BIBentryALTinterwordspacing}{\spaceskip=\fontdimen2\font plus
		\BIBentryALTinterwordstretchfactor\fontdimen3\font minus
		\fontdimen4\font\relax}
	\providecommand{\BIBforeignlanguage}[2]{{%
			\expandafter\ifx\csname l@#1\endcsname\relax
			\typeout{** WARNING: IEEEtran.bst: No hyphenation pattern has been}%
			\typeout{** loaded for the language `#1'. Using the pattern for}%
			\typeout{** the default language instead.}%
			\else
			\language=\csname l@#1\endcsname
			\fi
			#2}}
	\providecommand{\BIBdecl}{\relax}
	\BIBdecl
	
	\bibitem{dardari2020communicating}
	D.~{Dardari}, ``Communicating with large intelligent surfaces: Fundamental
	limits and models,'' \emph{IEEE J.\ Select.\ Areas \ Commun.}, vol.~38,
	no.~11, pp. 2526--2537, Nov. 2020.
	
	\bibitem{ConvergentCommunication}
	C.~{De Lima, \textit{et al.}}, ``Convergent communication, sensing and
	localization in {6G} systems: An overview of technologies, opportunities and
	challenges,'' \emph{IEEE Access}, vol.~9, pp. 26\,902--26\,925, 2021.
	
	\bibitem{wymeersch2019radio}
	H.~Wymeersch, J.~He, B.~Denis, A.~Clemente, and M.~Juntti, ``Radio localization
	and mapping with reconfigurable intelligent surfaces: Challenges,
	opportunities, and research directions,'' \emph{IEEE Vehicular Technology
		Magazine}, vol.~15, no.~4, pp. 52--61, Dec. 2020.
	
	\bibitem{Shahmansoori18TWC}
	A.~{Shahmansoori}, G.~E. {Garcia}, G.~{Destino}, G.~{Seco-Granados}, and
	H.~{Wymeersch}, ``Position and orientation estimation through millimeter-wave
	{MIMO} in {5G} systems,'' \emph{IEEE Trans.\ Wireless Commun.}, vol.~17,
	no.~3, pp. 1822--1835, Mar. 2018.
	
	\bibitem{eigenvector}
	J.~{Hong} and T.~{Ohtsuki}, ``Signal eigenvector-based device-free passive
	localization using array sensor,'' \emph{IEEE Trans.\ Vehicular Tech.},
	vol.~64, no.~4, pp. 1354--1363, Apr. 2015.
	
	\bibitem{rfidMag}
	L.~M. {Ni}, D.~{Zhang}, and M.~R. {Souryal}, ``{RFID}-based localization and
	tracking technologies,'' \emph{IEEE Wireless Commun.}, vol.~18, no.~2, pp.
	45--51, Apr. 2011.
	
	\bibitem{VehichlesOnRfid}
	H.~{Qin}, Y.~{Peng}, and W.~{Zhang}, ``Vehicles on {RFID}: Error-cognitive
	vehicle localization in {GPS}-less environments,'' \emph{IEEE Trans.\
		Vehicular Tech.}, vol.~66, no.~11, pp. 9943--9957, Nov. 2017.
	
	\bibitem{ruan2014tagtrack}
	W.~Ruan, L.~Yao, Q.~Z. Sheng, N.~J. Falkner, and X.~Li, ``Tagtrack: Device-free
	localization and tracking using passive {RFID} tags,'' in \emph{Proceedings
		of the 11th Int. Conf. on Mobile and Ubiquitous System}, London, UK, Dec.
	2014, pp. 80--89.
	
	\bibitem{shen2012accurate}
	J.~Shen, A.~F. Molisch, and J.~Salmi, ``Accurate passive location estimation
	using {TOA} measurements,'' \emph{IEEE Trans.\ Wireless Commun.}, vol.~11,
	no.~6, pp. 2182--2192, Jun. 2012.
	
	\bibitem{wang2014toa}
	Y.~Wang, S.~Ma, and C.~P. Chen, ``{TOA}-based passive localization in
	quasi-synchronous networks,'' \emph{IEEE Commun.\ Lett.}, vol.~18, no.~4, pp.
	592--595, Feb. 2014.
	
	\bibitem{ExpectationYuan}
	W.~{Yuan}, N.~{Wu}, B.~{Etzlinger}, Y.~{Li}, C.~{Yan}, and L.~{Hanzo},
	``Expectation–maximization-based passive localization relying on
	asynchronous receivers: Centralized versus distributed implementations,''
	\emph{IEEE Trans.\ Commun.}, vol.~67, no.~1, pp. 668--681, Jan. 2019.
	
	\bibitem{decarli2013novel}
	N.~{Decarli}, F.~{Guidi}, and D.~{Dardari}, ``A novel joint {RFID} and radar
	sensor network for passive localization: Design and performance bounds,''
	\emph{IEEE J.\ Select.\ Areas \ Commun.}, vol.~8, no.~1, pp. 80--95, Feb.
	2014.
	
	\bibitem{delayDoppler_passive_TSP_2016}
	X.~{Zhang}, H.~{Li}, J.~{Liu}, and B.~{Himed}, ``Joint delay and {Doppler}
	estimation for passive sensing with direct-path interference,'' \emph{IEEE
		Transactions on Signal Processing}, vol.~64, no.~3, pp. 630--640, 2016.
	
	\bibitem{keykhosravi2020siso}
	K.~Keykhosravi, M.~F. Keskin, G.~Seco-Granados, and H.~Wymeersch, ``{SISO}
	{RIS}-enabled joint {3D} downlink localization and synchronization,''
	accepted in \textit{IEEE Int. Conf. Commun. (ICC)}, Montreal, Canada, Jun.
	2021, available in arXiv preprint:2011.02391.
	
	\bibitem{fan2004spreading}
	P.~Fan, ``Spreading sequence design and theoretical limits for quasisynchronous
	{CDMA} systems,'' \emph{EURASIP J. on wireless Commun. and Net.}, vol. 2004,
	no.~1, pp. 1--13, Mar. 2004.
	
	\bibitem{kay1993fundamentals}
	S.~M. Kay, \emph{Fundamentals of statistical signal processing: Estimation
		Theory}.\hskip 1em plus 0.5em minus 0.4em\relax Prentice Hall PTR, 1993.
	
	\bibitem{malanowski2009algorithm}
	M.~Malanowski, ``An algorithm for {3D} target localization from passive radar
	measurements,'' in \emph{Photon. Appl. in Astron., Commun., Industry, and
		High-Energy Phys. Exp.}, Wilga, Poland, May 2009.
	
	\bibitem{ellingson2019path}
	S.~W. Ellingson, ``Path loss in reconfigurable intelligent surface-enabled
	channels,'' \emph{arXiv preprint arXiv:1912.06759}, 2019.
	
\end{thebibliography}
\end{document}